\documentclass[prd,nofootinbib,noshowpacs,superscriptaddress,preprint]{revtex4}
\usepackage[T1]{fontenc}
\usepackage{amsmath,amssymb}
\usepackage{epsfig}
\usepackage{dcolumn}
\usepackage{graphicx}
\usepackage[usenames]{color}
\usepackage{slashed}
\usepackage[colorlinks,citecolor=blue]{hyperref}
\usepackage{pdfpages}
\usepackage{float}
\usepackage{graphicx}
\usepackage{xcolor}
\let\revappendix\appendix

\def\bea{\begin{eqnarray}}
\def\eea{\end{eqnarray}}

\begin{document}
	
	\title{Neutrino masses and mixing in Minimal Inverse Seesaw using $A_4$ modular symmetry}
	\author{Jotin Gogoi}
	\email{jotingo@tezu.ernet.in}
	\author{Nayana Gautam}
	\email{nayana@tezu.ernet.in}
	\author{Mrinal Kumar Das}
	\email{mkdas@tezu.ernet.in}
	\affiliation{Department of Physics, Tezpur University, Tezpur 784028, India}
	\begin{abstract}
	In this paper, we construct a model with the help of modular symmetry in the framework of minimal inverse seesaw [ISS(2,3)]. We have used $\Gamma(3)$ modular group which is isomorphic to non-Abelian discrete symmetry group $A_4$. In this group there are three Yukawa modular forms of weight 2. Through this model, we study neutrino masses and mixing for both normal and inverted hierarchy. Use of modular symmetry reduces the need for extra flavons and their specific VEV alignments, as such, minimality of the model is maintained to a great extent. Along with $A_4$ symmetry group, we have used $Z_3$ to restrict certain interaction terms in the Lagrangian. Further we calculate the effective mass to address the phenomena of neutrinoless double-beta decay ($0\nu\beta\beta$). The values of effective mass is found to lie within the bound ($m_{eff}<0.165$ eV) as predicted by different $0\nu\beta\beta$ experiments.

	\end{abstract}
	\maketitle
	
\section{Introduction}	
	In the realm of particle physics, Standard Model is a well-established and experimentally verified model. It includes all known fundamental particles of matter present in the universe. The interactions among these fundamental particles along with the interaction-mediators, generation of masses through Higgs mechanism can be explained elaborately with the help of this model. With the discovery of Higgs boson in 2012 \cite{doi:10.1063/1.4727988}, both in ATLAS and CMS at the LHC this model was put on a firm footing, both experimentally and theoretically. As a result, predictions made from this model is in excellent agreement with that of experiments. There are three generations or flavors of leptons and quarks in Standard Model \cite{Ligeti:2015kwa,Xing:2020ijf}. All these generations of lepton and quark families together constitute the flavor part of particle physics. 
	
	In spite of robust successes, this model is not free from drawbacks. In absence of right-handed component of neutrino in the Standard Model, mass generation of the same could not be done in the model and so it remains massless. However with the discovery of neutrino oscillations \cite{Bilenky:1998dt,Bellini:2013wra}, it was confirmed that neutrinos should have mass, although it is very tiny. The discoveries of atmospheric and solar neutrino oscillations by Super-Kamioka Neutrino Detection Experiment (Super-Kamiokande) \cite{Super-Kamiokande:2001bfk} and Sudbury Neutrino Observatory \cite{SNO:2002hgz}, and thereafter confirmed by the Kamioka Liquid scintillator Antineutrino Detector (KamLAND) \cite{KamLAND:2008dgz} experiment provided sufficient evidence for neutrino flavor oscillation and massive neutrinos. Also solution to the problem of mass hierarchy \cite{Qian:2015waa,Ghosh:2012px} is yet to be found i.e. whether the neutrinos follow a normal mass spectrum ($m_1<m_2<m_3$) or  inverted mass spectrum ($m_3<m_1<m_2$). The exact nature of neutrinos (Dirac or Majorana) \cite{Barenboim:2002hx,Czakon:1999cd} is also an open question. Some of the other phenomena which could not be explained in the Standard Model are Baryon Asymmetry of the universe (BAU) \cite{Shaposhnikov:1987tw,Kolb:1979ui}, Dark matter \cite{Turner:1993dra,Taoso:2007qk}, dark energy \cite{Comelli:2003cv}, CP violation \cite{Ellis:1978hq} etc. Therefore to address and understand these enigmatic phenomena, we need new physics and improved experimental set-up to explore the domain beyond the reach of Standard Model.

	In recent years the problem of neutrino has gained much momentum and has been able to attract attention of the physics community at large. After the discovery of neutrino oscillations, followed by experimental evidences on neutrino masses and mixing,the need for new physics to provide a better understanding of the mentioned processes was of utmost importance. The breakthrough came through the Seesaw mechanism. It is one of the first mechanisms which could successfully explain generation of tiny neutrino masses. This mechanism is broadly classified into the following categories: Type I \cite{Mohapatra:1980yp,Minkowski:1977sc}, Type II \cite{Arhrib:2009mz}, Type III \cite{Ma:2002pf} and Inverse Seesaw (ISS) \cite{Mohapatra:1986aw,Deppisch:2004fa}. These mechanisms are basically an extension of SM with the addition of extra particles like right-handed neutrinos, scalar triplets, gauge singlet neutral fermions etc. There is ample sources of literature which make use of non-abelian discrete symmetry groups like $A_4$ \cite{Ma:2001dn,Babu:2002dz}, $S_4$ \cite{Altarelli:2009gn,Gautam:2019pce}, $A^{'}_{5}$ \cite{Feruglio:2011qq} etc. to study and understand the problem of flavor structure of fermions. These discrete symmetry groups are used to describe neutrino mixing phenomenology and neutrino mass matrix. But there are some drawbacks associated with this process of employing discrete symmetry groups to study neutrino phenomenologies.
	
	One of the primary disadvantages of discrete flavor symmetry is the inclusion of extra particles, called flavons to the model. These flavons are scalar multiplet fields whose vacuum expectation values have to be aligned in a particular way to fit experimental data. This paves the way for many unknown coefficients and non-renormalizable Yukawa interactions, as a consequence, predictability of the model comes under question. Because of these serious drawbacks, model building with the help of discrete flavor symmetry is often met with questions regarding the flavons and their VEV alignments.
		
	Modular symmetry is one of the aspects that has gained much importance in model building in recent times. Yukawa couplings in this framework is not free. In this approach \cite{King:2020qaj,Feruglio:2017spp} the Yukawa couplings and neutrino masses are expressed in terms of modular forms which are functions of a complex variable, called modulus and represented by $\tau$. The only source of symmetry breaking in this process is the complex variable $\tau$ and so there is minimum or no requirement of flavons in the model. This special character of modular symmetry keeps minimality of model intact. The modular group $\Gamma(N)$ acts on the upper half of the complex plane (Im($\tau)>$  0) as linear fractional transformations: $$\tau \rightarrow \frac{a\tau + b}{c\tau +d} $$
	where $a,b,c,d$ are integers and $ad-bc=1$.
	A lot of work in different frameworks has been done using modular symmetry to understand and realize the processes of neutrino mass generation and mixings.  For instance, in the literatures \cite{Novichkov:2019sqv,Wang:2019ovr} seesaw mechanisms, Type II \cite{Kobayashi:2019gtp,Kashav:2021zir}, inverse seesaw \cite{Nomura:2019xsb,Zhang:2021olk} , scotogenic \cite{Nomura:2019lnr}, fermion mass hierarchies etc. have been explored with the help of modular symmetry.
	
	The nature of neutrinos is still a challenging problem in particle physics: whether the neutrinos are of Dirac or Majorana type is not known to us. Neutrinoless double-beta decay (NDBD/$0\nu\beta\beta$) is one of the attempts to explain and address this problem \cite{Klapdor-Kleingrothaus:2001oba,Pas:2015eia}. It is a process in which lepton number is violated by two units ($\Delta L=2$) and can be described by the reaction: $$(A,Z)\rightarrow(A,Z+2)+2e^-$$ Currently there are ongoing experiments, such as KamLANDZen \cite{KamLAND-Zen:2016pfg} and GERDA \cite{GERDA:2020xhi} which use Xenon-136 and Germanium-76 nuclei respectively, to detect this lepton number violating reactions, which if detected once, will confirm Majorana nature of neutrinos and also shade light with respect to absolute scale of neutrino masses. Measurement of the effective Majorana neutrino mass, which is a combination of neutrino mass eigenstates and mixing matrix terms, is one of the main objectives of NDBD. The results published from KamLAND-Zen experiments predicts the effective mass to lie in the range (0.061-0.165) eV. Moreover the lower bound for life-time of the decay is found to be $T^{0\nu}_{1/2}$\textgreater1.07$\times 10^{26}$ yrs. In this work we try to study the extended contributions to NDBD processes using modular symmetry.
	
	This paper is arranged in the following way: in section \ref{sec:level2} we briefly discuss about the Inverse seesaw framework and then continue to explain the mechanism of Minimal Inverse seesaw[ISS(2,3)]. The constructed model is discussed in section \ref{sec:level3}. In section \ref{sec:level4},we discuss about the NDBD process in the framework of ISS(2,3). In section \ref{sec:level5}, we provide the results and numerical analysis of this work and finally in section \ref{sec:level6} we conclude by giving an overview of our work.
	
	\section{\label{sec:level2}Inverse Seesaw: [ISS(2,3)] framework}
	In the literature \cite{King:2003jb,Mohapatra:2005wg,King:2014nza}, there are different seesaw mechanisms which try to explain the processes of neutrino mass generation. These mechanisms are extensions of the Standard Model with the addition of extra particles like right-handed neutrinos, scalar triplets etc. In order to explain the smallness of active neutrino masses in natural order of Yukawa couplings, masses of the extra particles need to be of very high order scale. For instance, in the conventional seesaw mechanism mass of the right-handed neutrinos is of the order of $10^{15}$ GeV, making it quite difficult for detection. On the contrary, Inverse seesaw is a quite realistic and experimentally probable approach that aims to explain generation of tiny neutrino masses in a much feasible way \cite{Deppisch:2004fa,Dev:2009aw}. It is also an extension of SM with the addition of $SU(2)\otimes U(1)$ singlet right-handed neutrinos ($N_{Ri}$), just like the conventional Type I seesaw. Along with these particles, it has neutral gauge singlet sterile fermions ($S_i$). This framework has the advantage that it can lower the mass-scale of right-handed neutrinos to TeV, due to which there is scope of them being detected in the LHC and future neutrino oscillation experiments. As a result, compared to the other conventional mechanisms, Inverse seesaw becomes a more realistic and experimentally feasible framework. From the literature \cite{Abada:2014zra,Abada:2014vea,Abada:2017ieq}, the lagrangian of the model can be written as:
	\begin{equation}
	L=-\frac{1}{2}n^T_LCMn_L+h.c.
	\label{eqn:1}
	\end{equation}
	where $C\equiv \gamma^2 \gamma^0$ is the charge conjugation matrix and the basis $n_L=(\nu_{L,\alpha},N^c_{R,i},S_j)^T.$ The component $\nu_{L,\alpha}$ for $\alpha=e,\mu,\tau$ are left-handed Standard Model neutrinos. The complete mass matrix for neutral fermion arising from the lagrangian can be written as:
	\begin{equation}
	M= \begin{pmatrix}
	0&M^T_D&0\\M_D&0&M_{NS}\\0&M^T_{NS}&M_S
	\end{pmatrix}
	\label{eqn:2}
	\end{equation} 
	Here $M_D$ is the Dirac mass matrix resulting from interaction between the left and right-handed components of neutrino. Interaction between the right-handed neutrinos and sterile singlet fermions, commonly referred to as the mixing term, is represented by $M_{NS}$. For the majorana interaction between the sterile singlet neutral fermions, referred to as Majorana mass term ,is represented by $M_S$.

	The matrix $M$ can be diagonalised with the help of a Unitary matrix, $\mathcal{U}$ as
	\begin{equation}
	\mathcal{U^T}M\mathcal{U}=M_{diag}=diag(m_1,m_2,m_3,....,m_8)
	\label{eqn:3}
	\end{equation}
	where $m_i$'s are masses of the particles of the model.

	The effective neutrino mass matrix of the active light neutrinos can be written as: \begin{equation}
	m_\nu=M^T_D(M^T_{NS})^{-1}M_SM^{-1}_{NS}M_D
	\label{eqn:4}
	\end{equation}
	In order to produce sub-eV Standard Model neutrinos, $M_D$ must be in electroweak range, $M_{NS}$ in the TeV range and $M_S$ must be in the KeV range, respectively. Dimensions of the  matrices are associated with the number of generations of the particles that are considered in the model. Accordingly, for these matrices, the dimensions can be defined in the following way (\# represents the number of generations of particles.):
	$$Dimension \hspace{1.5mm} of\hspace{1.5mm} M_D=(\#\nu_L\times \#N_R)$$
	
	$$Dimension \hspace{1.5mm} of\hspace{1.5mm} M_{NS}=(\#N_R\times \#S)$$

	$$Dimension \hspace{1.5mm} of\hspace{1.5mm} M_S=(\#S\times \#S)$$
	
	In our work we have used the Minimal Inverse Seesaw [ISS(2,3)] framework \cite{Abada:2014zra,Abada:2014vea}. This mechanism of neutrino mass generation involves two right-handed neutrinos, along with three gauge singlet neutral fermions. The mass matrix in (\ref{eqn:2}) can be decomposed into light and heavy sectors where diagonalisation of light sector gives the masses of light neutrinos and heavy sector is responsible for mass eigen values of the other five particles. Light sector can be expressed as equation (\ref{eqn:4}), whereas matrix for the heavy sector is:
	\begin{equation}
	M_H=\begin{pmatrix}
	0&M_{NS}\\M^T_{NS}&M_S
	\end{pmatrix}
	\label{eqn:5}
	\end{equation}
	From the dimensions one can find that $M_{NS}$ is not a square matrix, so its inverse is not properly defined. As a result, the effective neutrino mass matrix changes slightly and takes a different form as compared to (\ref{eqn:4}). In order to retain a similar pattern as that of (\ref{eqn:4}), the mass matrix can be written as \cite{Abada:2017ieq}:
	\begin{equation}
	m_\nu=M_D.d.M^T_D
	\label{eqn:6}
	\end{equation}
	where $d$ is a 2$\times$ 2 matrix and it is defined as:
	\begin{equation}
	M^{-1}_H=\begin{pmatrix}
	d_{2\times 2}&......\\......&......
	\end{pmatrix}
	\label{eqn:7}
	\end{equation}
	Now diagonalising equation (\ref{eqn:6}), one can get the masses of light active neutrinos.

	
	\section{\label{sec:level3}The Model}
	
	Here, in this section, we will discuss about the model and then construct the associated Lagrangian with the help of $A_4$ modular symmetry. The particle content of the model is an extension of the Standard Model (SM) with a pair of $SU(2)\otimes U(1)$ singlet right-handed neutrinos, $N_i (i=1,2)$. In addition to these, there are three neutral gauge singlet fermions, $S_i (i=1,2,3)$. Moreover, $Z_3$ symmetry is used to get the desired interacting terms in the Lagrangian.\vspace{2mm}
	
	The non-Abelian discrete symmetry group, $A_4$ has four irreducible representations \cite{Ishimori:2010au}: a triplet and three singlets. The triplet is represented as $3$ and the singlets are represented as $(1,1',1'')$. Product rules of $A_4$ is given in Appendix \textcolor{red}{B}. The lepton doublets (L) transform as $A_4$ triplets and right handed charged leptons $E_i (i=e,\mu,\tau)$ transforms as $1,1''$ and $1'$, respectively. The right-handed neutrinos $N_1$ and $N_2$ transforms as $1'$ and $1''$ ; whereas the Higgs doublets $(H_d,H_u)$ transform as trivial singlet (1) of the discrete non-abelian group $A_4$. The neutral singlet gauge fermions $S_i (i=1,2,3)$ are assigned triplet representation of the $A_4$ group. Apart from these particles, a flavon $\phi$ is used in the model to get a diagonal mass matrix for the charged leptons. Modular weights of lepton doublets is zero whereas that of the right-handed charged leptons is -2. Similarly the right-handed neutrinos are assigned modular weights of -2 and rest of the particles ($S_i,H_u,H_d$) are taken to be of zero modular weights. The flavon is assigned a modular weight of 2. The charge assignments in different symmetry groups and modular weights of the particles are shown in the following Table (\ref{tab:A}).  \vspace{2mm}

	\begin{table}[ht]
		\centering
		\begin{tabular}{|c|c|c|c|c|c|c|c|c|}
			\hline
			&L & $E_i$ & $N_1$ &$N_2$ & $S_i$ &$H_u$ &$H_d$&$\phi$\\
			\hline
			$SU(2)_L$ &2 & 1 & 1 & 1 & 1 & 2 & 2 &1\\
			\hline
			$A_4$ & 3 &$1,1'',1'$ &$1'$ & $1''$ &3 &1 &1&3\\
			\hline
			$K_I$ & 0 & -2 & -2 &-2 & 0 &0 &0&2\\
			\hline
			$Z_{3}$ & $\omega^2$ & 1& $\omega$ & $\omega$ & 1 &1 & $\omega$ & $\omega$\\
			\hline
			
		\end{tabular}
		\caption[]{The above table shows the charge assignments and modular weights of the particles considered in the model. $K_I$ denotes modular weights of the particles.}
		\label{tab:A}
	\end{table}

	The Yukawa couplings, $Y=(Y_1,Y_2,Y_3)^T$, are expressed as modular forms of level three and weight 2 of the modular group $\Gamma(3)$. A brief description of modular symmetry is given in Appendix \textcolor{red}{A}. They are three in number and so are expressed as a triplet of $A_4$.
	
	\begin{table}[ht]
		\centering
		\begin{tabular}{|c|c|}
			\hline
			&Y (Modular forms)\\
			\hline
			$A_4$& 3\\
			\hline
			$K_I$ & 2\\
			\hline
			$Z_3$&$\omega^2$\\
			\hline
		\end{tabular}
		\caption{The charge assignments and weight of Yukawa modular forms for the corresponding groups are shown in the above table.}
	\end{table}
	
	Based on the above discussions on charge assignments and symmetries, the Lagrangian of the model for the leptonic sector can be written in the following way:
	\begin{equation}
	-L= L_L+L_D + L_{NS} + L_S
	\label{eqn:8}
	\end{equation}
	where $L_L$ is the mass term for charged leptons, $L_D$ is the Dirac mass term connecting left-handed ($\nu_L$) and right-handed ($N_R$) components of neutrinos. $L_{NS}$ represents the mixing term between right-handed neutrinos and sterile fermions  ($S_i$) and $L_S$ is the majorana mass term among gauge singlet sterile fermions ($S_i$). All the terms of the Lagrangian must be invariant under $A_4$ symmetry group and sum of the modular weights of each term must be zero. Here we denote the vacuum expectation value of $H_d$ and $H_u$ as $\big\langle H\big\rangle=v$.\vspace{3
		mm}

	The VEV of the flavon is chosen in the following way \cite{Feruglio:2017spp}:
	\begin{equation}
	\big\langle\phi\big\rangle=(u,0,0).
	\label{eqn:9}
	\end{equation}
	
	The lagrangian for charged leptons take the form:
	\begin{equation}
	L_L=\alpha_1E^c_1H_d(L\phi)_1+ \alpha_2 E^c_2H_d(L\phi)_{1'}+ \alpha_3E^c_3H_d(L\phi)_{1''}
	\label{eqn:10}
	\end{equation}
	The parameters $\alpha_1,\alpha_2,\alpha_3$ can be adjusted to get the desired charged lepton masses. With the VEV of the flavon mentioned above, the diagonal mass matrix is obtained as:
	\begin{equation}
	M_L=diag(\alpha_1,\alpha_2,\alpha_3)u \hspace{2mm}v
	\label{eqn:11}
	\end{equation}
	
	Following the above discussions, the relevant Dirac mass term  of  the model can be written as follows :
	\begin{equation}
	L_D= [N_{1}(LY)_{1''}H_u]_1+[N_{2}(LY)_{1'}H_u]_1
	\label{eqn:12}
	\end{equation}
	
	where subscripts ($1',1'',1$) represents the irreducible representations of $A_4$ symmetry group. Subsequently the Dirac mass matrix for the neutrinos is obtained from equation (\ref{eqn:12}) in the following form:
	\begin{equation}
	M_D= \hspace{1.5mm} v \begin{pmatrix}
	Y_3 & Y_2\\Y_2 & Y_1 \\ Y_1 & Y_3 
	\end{pmatrix}
	\label{eqn:13}
	\end{equation}
	
	The Majorana mass term for the gauge singlet fermion is:
	\begin{equation}
	L_S= \Lambda\hspace{1.5mm} (SS)_1
	\label{eqn:14}
	\end{equation}
	
	The mass matrix from equation (\ref{eqn:14}) can be written as:
	\begin{equation}
	M_S= \hspace{1.5mm} \Lambda \begin{pmatrix}
	1 & 0&0\\0 & 0&1 \\ 0 & 1&0 
	\end{pmatrix}
	\label{eqn:15}
	\end{equation}
	
	Similarly, the mixing mass term between the right-handed neutrinos and gauge singlet neutral fermions can be expressed as:
	\begin{equation}
	L_{NS} =\hspace{1.5mm}\beta \hspace{1.5mm} [\{N_{1}(SY)_{1''}\}+\{N_{2}(SY)_{1'}\}]
	\label{eqn:16}
	\end{equation}
	In equations (\ref{eqn:14}) and (\ref{eqn:16}) , $\Lambda$ and $\beta$ are free parameters. From this lagrangian mass matrix  for the mixing term is obtained as:
	\begin{equation}
	M_{NS}= \hspace{1.5mm} \beta \begin{pmatrix}
	Y_3 & Y_2& Y_1\\Y_2 & Y_1  & Y_3 
	\end{pmatrix}
	\label{eqn:17}
	\end{equation}
	Taking all the mass terms together, we can express the entire Lagrangian in a single $8\times8$ neutrino mass matrix in the basis $(\nu_i, N_j, S_i)^T$, where $(i=1,2,3)$ and $(j=1,2)$. The eigenvalues of this matrix correspond to the mass of the eight fermions, respectively.  Finally this single neutrino matrix, in terms of $M_D$, $M_{NS}$ and $M_S$ can be written as:
	\begin{equation}
	M_\nu=\begin{pmatrix}
	0& M_D & 0\\ {M_D}^T & 0 & M_{NS}\\ 0 & {M_{NS}}^T& M_S
	
	\label{eqn:18}
	\end{pmatrix}
	\end{equation}
	
	After block diagonalising $M_\nu$, the active light neutrino mass matrix can be written from equation (\ref{eqn:6}) in the following way:
	\begin{equation}
	m_\nu=M_D\hspace{1mm}.\hspace{1mm} d\hspace{1mm}.\hspace{1mm}{M_D}^T
	\label{eqn:19}
	\end{equation}
	
	In this way, in the framework of ISS(2,3), we built a model using modular symmetry. Along with $A_4$ symmetry group, to constrain and restrict certain interaction terms in the Lagrangian, $Z_3$ is used. Using this model we carry out further works to obtain neutrino parameters and the phenomena of neutrinoless double-beta decay.
	

	\section{\label{sec:level4}Neutrinoless Double Beta Decay }
	It has been mentioned that ISS (2,3) contains eight extra heavy states and these states may have significant contributions to lepton number violating processes like neutrinoless double beta decay(0$\nu\beta\beta$) \cite{Benes:2005hn,Awasthi:2013we}. We have studied the effective electron neutrino Majorana mass $m_{ee}$ \cite{Abada:2018qok,Blennow:2010th} characterising 0$\nu\beta\beta$ in this model. Experiments like KamLAND-ZEN, GERDA, CUORE and EX0-200 provide stringent bounds on $m_{ee}$ which can be seen in  \cite{KamLAND-Zen:2016pfg,CUORE:2019yfd,GERDA:2020xhi}.
	
	The decay width of the process is proportional to the effective electron neutrino Majorana mass $m_{ee}$. In the absence of any sterile neutrino,the standard contribution to $m_{ee}$ can be written as,
	\begin{equation}
	m_{ee} = \mathrel{\Big|}\sum_{i = 1}^{3}{U_{ei}}^{2}m_{i}\mathrel{\Big|}
	\end{equation}
	The above equation is modified because of the presence of the heavy neutrinos and is given by \cite{Abada:2018qok} 
	\begin{equation}\label{eq:30}
	m_{ee}  =\mathrel{\Big|}\sum_{i = 1}^{3}{U_{ei}}^{2}m_{i}\mathrel{\Big|} + \mathrel{\Big|}\sum_{j = 1}^{5}{U_{ej}}^{2}\frac{M_{j}}{k^{2}+M_{j}^{2}}|<k>|^{2}\mathrel{\Big|}
	\end{equation}
	where, ${U_{ej}}$ represents the coupling of the heavy neutrinos to the electron neutrino and $M_{j}$ represents the mass of the respective heavy neutrinos. $|<k>|$ is known as neutrino virtuality momentum with value $|<k>|\simeq 190$ MeV.
	
	\section{\label{sec:level5}Numerical analysis and results}
	
	To carry out the numerical analysis, we have considered the following 3$\sigma$ experimental values of neutrino oscillation parameters \cite{Esteban:2020cvm}, summarised in table \ref{tab:III}:
	
	\begin{table}[ht]
		\centering
		\begin{tabular}{||c|c|c||}
			\hline
			Parameters & Normal Ordering & Inverted Ordering\\
			\hline
			$\sin^2{\theta_{12}}$ & [0.269,0.343] & [0.269,0.343] \\
			\hline
			$\sin^2{\theta_{23}}$ & [0.407,0.618] & [0.411.0.621] \\
			\hline
			$\sin^2{\theta_{13}}$ & [0.02034,0.02430] & [0.02053,0.02436] \\
			\hline
				$\Delta m^2_{21}/10^{-5}eV^2$ & [6.82,8.04] & [6.82,8.04] \\
			\hline
			$\Delta m^2_{31}/10^{-3}eV^2$ & [2.431,2.598] & [2.412,2.583] \\
			\hline
		\end{tabular}
	\caption{The table above shows the 3$\sigma$ values of neutrino oscillation parameters.}
	\label{tab:III}
	\end{table}

Here we numerically diagonalise the neutrino mass matrix using the relation, $m_\nu=U .\hspace{2mm}m_{diag}.\hspace{2mm} U^T$, where $U$ is a $3\times3$ unitary matrix and $m_{diag}$= diag$(m_1,m_2,m_3).$ The mixing angles can be obtained from the elements of unitary matrix $U$ in the following way \cite{Behera:2021eut}:
\begin{equation}
\sin^2{\theta_{13}}=|U_{e3}|^2,
\hspace{5mm} 
\sin^2{\theta_{23}}=\frac{|U_{\mu 3}|^2}{1-|U_{e3}|^2},
\hspace{5mm} \sin^2{\theta_{12}}=\frac{|U_{e2}|^2}{1-|U_{e3}|^2}
\label{eqn:20}
\end{equation}

Moreover, the Jarlskog invariant that controls the size of CP violation in quark and lepton sector can be calculated from the elements of the matrix $U$ in the following way:
\begin{equation}
J_{cp}= Im[U_{e1}U_{\mu 2}U^*_{e2}U^*_{\mu 1}]=s_{23}c_{23}s_{12}c_{12}s_{13}c^2_{13}\sin\delta_{cp}
\label{eqn:21}
\end{equation}

	In order to fit the neutrino oscillation data, we have taken the following ranges of values for the model parameters:
	
	\vspace{2mm}
	Re$(\tau)\rightarrow[0,3]$, \hspace{5mm}Im$(\tau)\rightarrow[0.04,2.8]$,\hspace{5mm} $v=125 \hspace{1mm}GeV$, 
	
	\vspace{2mm}
	$\Lambda\rightarrow[10,20] KeV$, \hspace{5mm}$\beta\rightarrow[10,100] TeV$

	\vspace{5mm}
	With the help of the $U_{PMNS}$ matrix, we diagonalise the light neutrino mass matrix for both normal and inverted ordering of neutrino parameters. The solar and atmospheric mass squared differences provide necessary constraints for the model. The values of $\Lambda$ in the range (10 - 20) KeV, apart from producing neutrino phenomenology, will be helpful to explore and study sterile neutrino dark matter. Using $q$-expansions of the modular forms, the real and imaginary parts of complex modulus $\tau$ is found to lie in the following range: Re($\tau)\rightarrow[0,3]$ and Im($\tau)\rightarrow[0.04,2.8]$. The mixing angles from the model can be obtained using the relations given in equation (\ref{eqn:20}). Using the model parameters, we have calculated the effective masses in case of normal and inverted hierarchy using equation (\ref{eq:30}).

\begin{figure}[H]
		\includegraphics[scale=0.3]{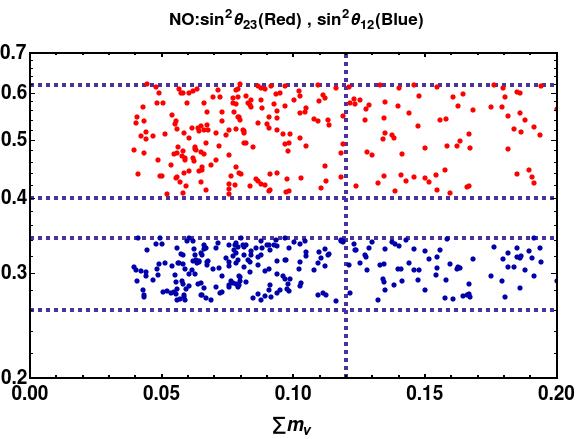}
		\hspace{7mm}
		\includegraphics[scale=0.3]{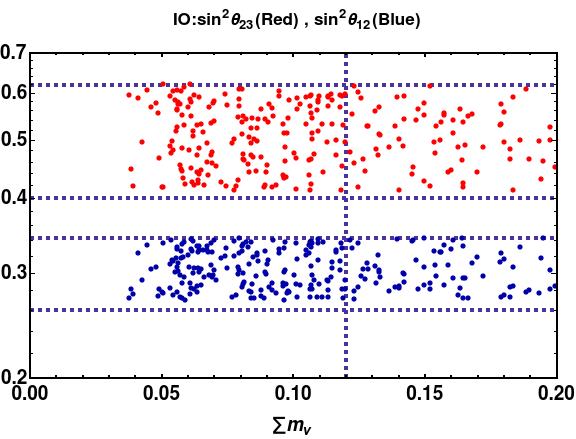}
		
		\caption{The correlation between sum of neutrino masses($\sum m_\nu$) and mixing angles, $\sin^2\theta_{23}$ and $\sin^2\theta_{12}$ for NO (left) and for IO(right).}
		\label{fig:1}
	\end{figure}

\begin{figure}[H]
		\includegraphics[scale=0.3]{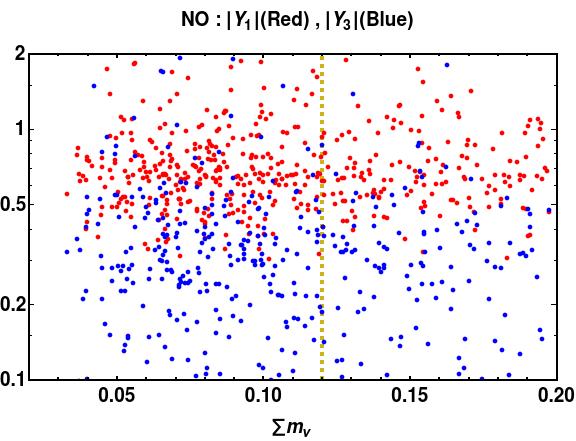}
		\hspace{7mm}
		\includegraphics[scale=0.3]{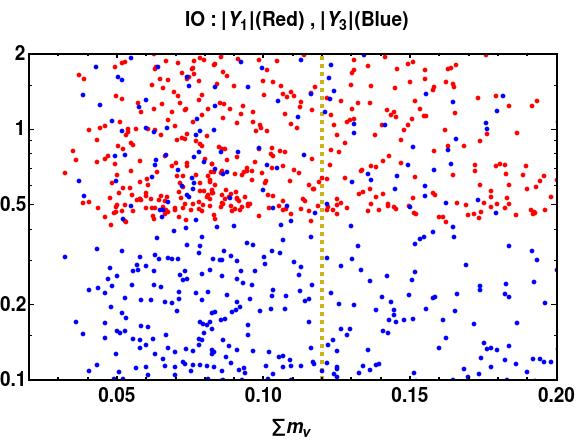}
		\caption{Correlation between yukawa couplings ($|Y_1|$ and $|Y_3|$) and $\sum m_\nu$ for NO (left) and for IO(right).}
		\label{fig:8}
	\end{figure}

	We have shown the results of our numerical analysis from fig \ref{fig:1} to fig \ref{fig:8}. 

	\begin{figure}[H]
		\includegraphics[scale=0.3]{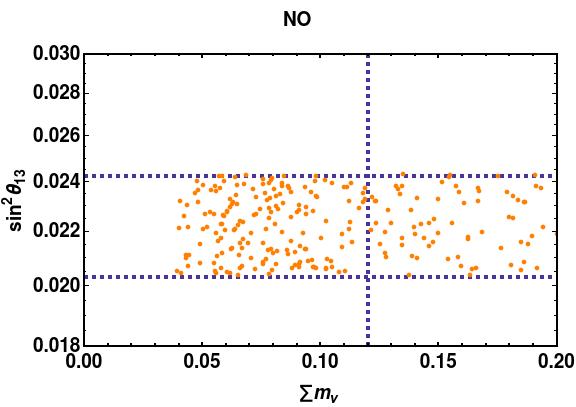}
		\hspace{7mm}
		\includegraphics[scale=0.3]{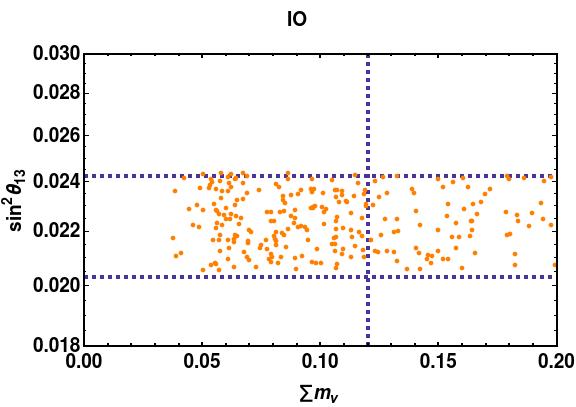}
		\caption{The variation of mixing angle, $\sin^2\theta_{13}$ as a function of sum of neutrino masses ($\sum m_{\nu}$) for NO (left) and for IO(right). }
		
		\label{fig:2}
		
	\end{figure}
	
	In figures (\ref{fig:1}) and (\ref{fig:2}), we show the variation between sum of neutrino masses ($\sum m_\nu$) and mixing angles $\sin^2\theta_{12}$/ $\sin^2\theta_{23}$ and $\sin^2\theta_{13}$ respectively. It is evident that a wide range of parameter space lie within the allowed region for upper bound of sum of neutrino masses ($\sum m_\nu < 0.12 eV$) and also the mixing angles. The lower bound for sum of neutrino masses is found to be around 0.05 $eV$ for both the cases.
	
\begin{figure}[H]
		\includegraphics[scale=0.3]{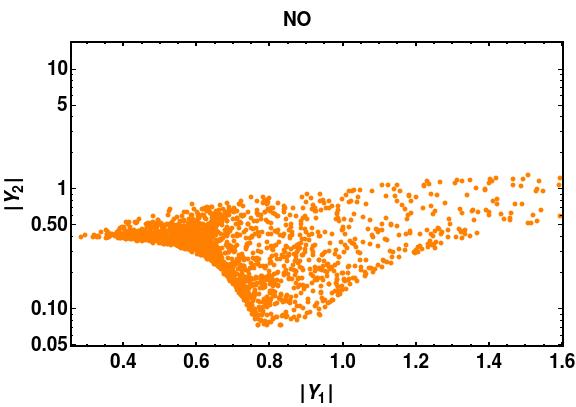}
		\hspace{7mm}
		\includegraphics[scale=0.3]{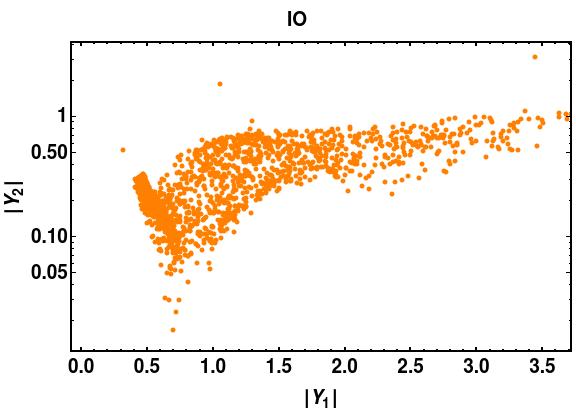}
		\caption{Variation between the Yukawa couplings $|Y_1|$ and $|Y_2|$ for both normal and inverted ordering.}
		\label{fig:3}
	\end{figure}

\begin{figure}[H]
		\includegraphics[scale=0.3]{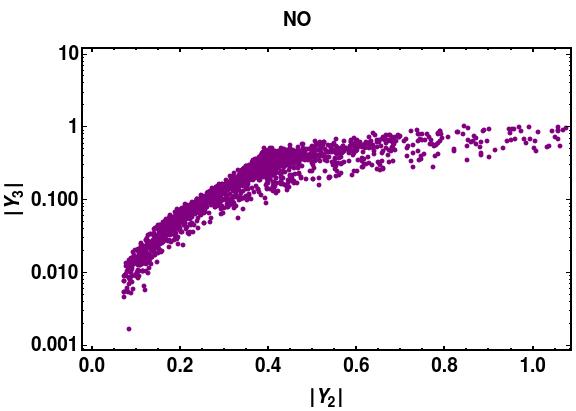}
		\hspace{7mm}
		\includegraphics[scale=0.3]{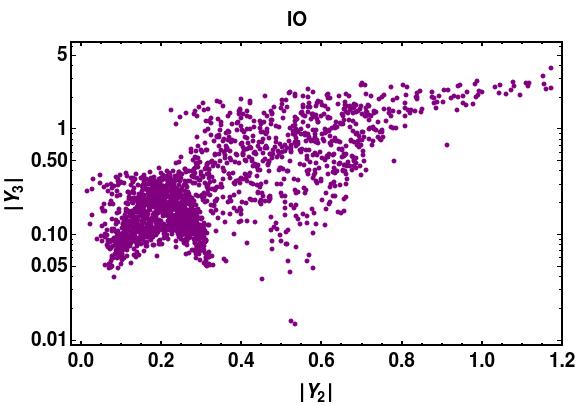}
		\caption{The figures show variation between the yukawa couplings $|Y_2|$ and $|Y_3|$ for both normal and inverted ordering.}
		\label{fig:4}
	\end{figure}

	Fig. (\ref{fig:3}) and (\ref{fig:4}) represent the variation between the Yukawa modular forms. It has been observed from fig. (\ref{fig:3}) that the value of $|Y_1|$ in case of NO lie within (0.1-1.6) and $|Y_2|$ lie in the region (0.07-1). While for the IO case, $|Y_1|$ mostly lie in the region (0.5-3) and $|Y_2|$ lie in the region (0.01-1). Similarly in fig. (\ref{fig:4}), we see that $|Y_3|$ lie within the region (0.009-0.99) and (0.05-1.99) for NO and IO respectively. These values of the yukawa couplings have been summarised below in table \ref{tab:IV}:
	
	\begin{table}[ht]
		\centering
		\begin{tabular}{||c|c|c||}
			\hline
			& Normal Ordering & Inverted Ordering\\
			\hline
			|$Y_1$| &  0.1 - 1.6      &  0.5 - 3   \\
			\hline
			|$Y_2$| &   0.07 - 1     &  0.01 - 1    \\
			\hline
			|$Y_3$| &   0.009 - 0.99     & 0.05 - 1.99     \\
			\hline
		\end{tabular}
	\caption{Values (range) of the Yukawa couplings obtained from the model.}
	\label{tab:IV}
	\end{table}
	
\begin{figure}[H]
		\includegraphics[scale=0.3]{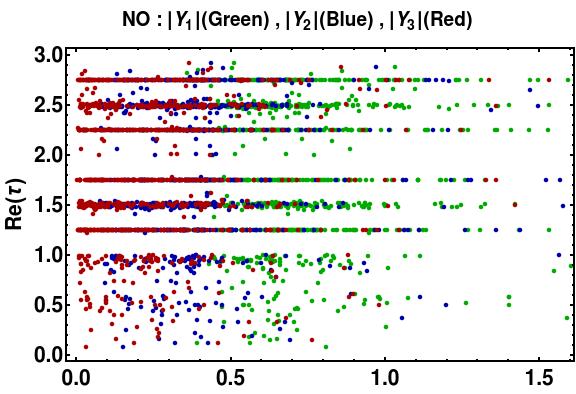}
		\hspace{7mm}
		\includegraphics[scale=0.3]{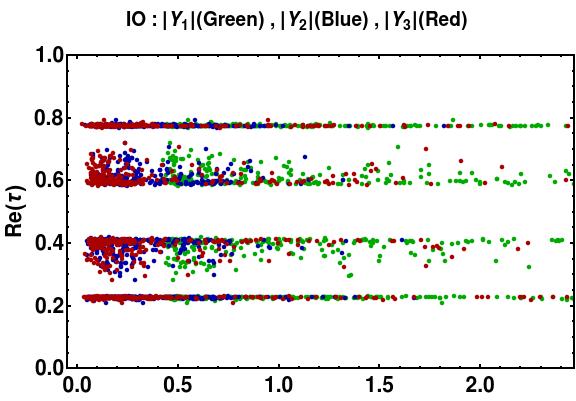}
		\caption{Variation of the Yukawa couplings as a function of the real part of $\tau$ for NO (left) and for IO(right).}
		\label{fig:5}
	\end{figure}

\begin{figure}[H]
		\includegraphics[scale=0.3]{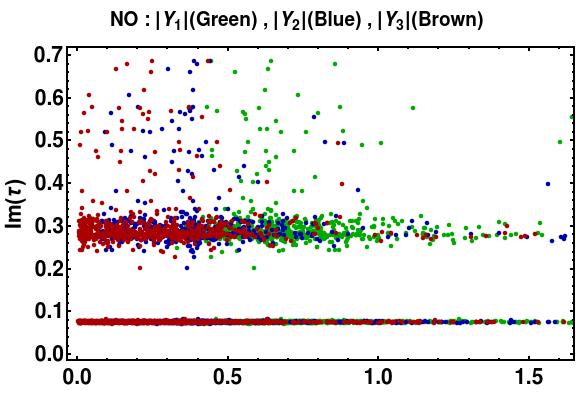}
		\hspace{7mm}
		\includegraphics[scale=0.3]{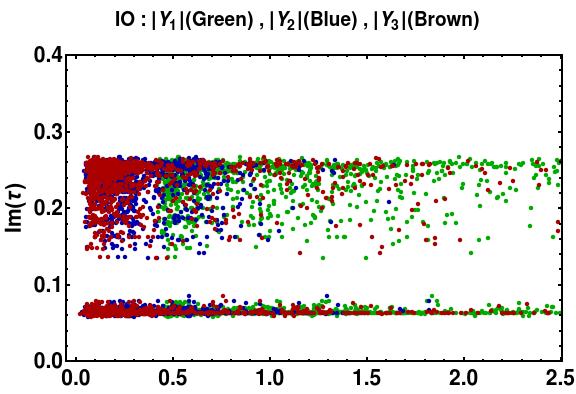}
		\caption{Variation of the Yukawa couplings as a function of the imaginary part of $\tau$ for NO (left) and for IO(right).}
		\label{fig:6}
	\end{figure}

	Fig (\ref{fig:5}) shows the variation between Yukawa couplings and real part of the complex modulus, $\tau$ for both normal and inverted ordering. There are certain regions of Re($\tau$)  where the parameter space of the Yukawa couplings lie. Similarly, the variations of imaginary part of $\tau$ with respect to the Yukawa couplings are shown in fig. (\ref{fig:6}). Here we find that Yukawa couplings are confined to two regions of Im($\tau$). For NO this regions lie near (0.08-0.1) and (0.25-0.31), while for IO it is found near (0.06-0.09) and (0.15-0.27).

	\begin{figure}[H]
		\includegraphics[scale=0.3]{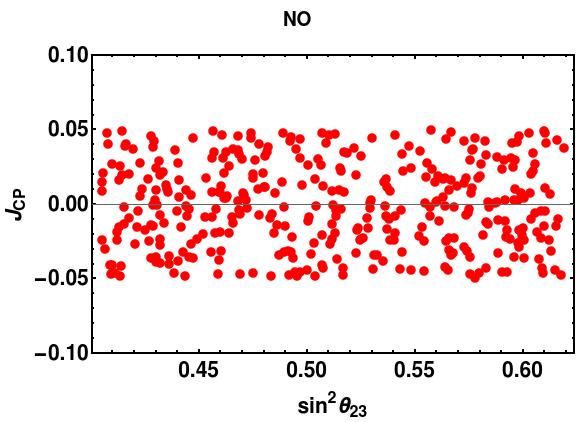}
		\hspace{7mm}
		\includegraphics[scale=0.3]{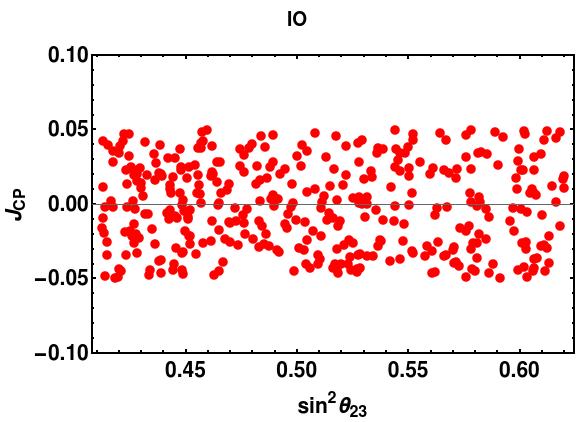}
		\caption{Correlation between $J_{cp}$ and $\sin^2\theta_{23}$ for NO (left) and for IO(right).}
		\label{fig:7}
	\end{figure}
	
	The correlation between Jarlskog invariant and $\sin^2\theta_{23}$ is shown in fig. (\ref{fig:7}). For both the ordering, there are sufficient values of $J_{cp}$ which lie within the range (-0.04-0.04). 
	
	Finally we show the variations between Yukawa couplings, $|Y_1|$ and $|Y_3|$ with sum of neutrino masses ($\sum m_\nu$) in fig. (\ref{fig:8}). For normal ordering, most of the values of $|Y_1|$ is found to lie in the region (0.5-1) and that of $|Y_3|$ is scattered towards the lower part  with points reducing rapidly after 0.2. While for inverted ordering, the Yukawa coupling |$Y_1$| is mostly confined to the region above 0.5 and that of $|Y_2|$ is mainly spanned in the region (0.1-2).
	
		\begin{figure}[H]
		\includegraphics[scale=0.3]{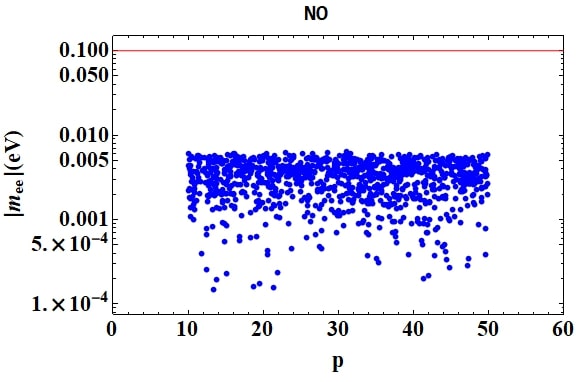}
		\hspace{7mm}
		\includegraphics[scale=0.3]{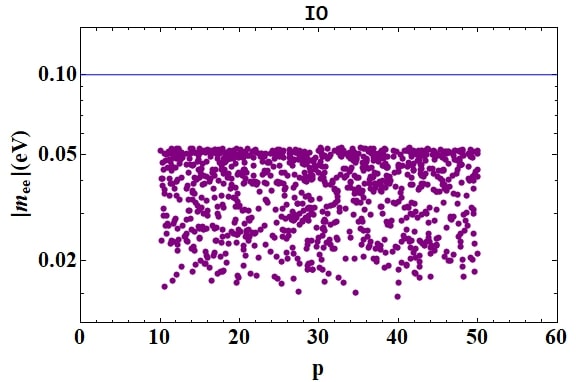}
		\caption{Effective mass as a function of parameter p for  NO (left) and for IO(right).}
		\label{fig:9}
	\end{figure}
	Fig (\ref{fig:9}) depicts the prediction of the model on effective mass characterizing NDBD process for both NO and IO. It is observed that the effective mass lie within the experimental limits in both the cases.
	
	
	\section{\label{sec:level6} Conclusion}
In this work we have constructed a model with the help of modular symmetry in the framework of minimal inverse seesaw, ISS(2,3). This framework includes two right-handed neutrinos and three gauge singlet neutral fermions. We have used $\Gamma(3)$ (isomorphic to $A_4$) symmetric group to construct the relevant Lagrangian of the model and then obtain the different corresponding mass matrices. Certain interactions in the Lagrangian have been constrained with the appropriate use of $Z_3$. In this model, the number of flavons is significantly reduced compared to the previous work with $S_4$ flavor symmetry \cite{Gautam:2019pce}, as we have used only one flavon in this model. The construction of the model in ISS(2,3) with $A_4$ modular symmetry has not been done before. Proceeding ahead, we evaluate the model parameters and subsequently study the prediction of the model on different neutrino parameters. A sufficient space of the sum of active neutrino mass ($\sum m_\nu$) is found to lie within the upper bound i.e. 0.12 eV, whereas the lower value is found to lie around 0.03 eV. The Yukawa couplings are constrained and found to lie mostly in the region (0.01-2). We  have also studied NDBD in the model and have found that the model is consistent with the current experimental bounds. One can study dark matter and also baryon asymmetry of the universe within the model that we leave for our future study.

\revappendix
\section{Modular Symmetry}

The modular group $\Gamma(N)$ ($N=1,2,3,....)$ can be defined as\begin{equation}
\Gamma(N)=\{\begin{pmatrix} a & b\\c & d \end{pmatrix} \in SL(2,Z) , \begin{pmatrix} a & b\\c & d \end{pmatrix}=\begin{pmatrix} 1 & 0\\0 & 1 \end{pmatrix} (modN)\}
\label{eqn:A1}
\end{equation} such that $ad-bc=1.$

The modular group $\Gamma(N)$ acts on complex variable $\tau$ in the upper-half of the complex plane (Im$\tau$ \textgreater \hspace{2mm}0) and transforms $\tau$ in the following way:$$\tau \rightarrow \frac{a\tau + b}{c\tau+d}.$$The modular group has two generators:
\begin{align}
S=\begin{pmatrix} 0 & 1\\-1 & 0\end{pmatrix} ,  & \hspace{2cm}  T=\begin{pmatrix} 1 & 1\\0 & 1\end{pmatrix}
\label{eqn:A2}
\end{align} such that

\begin{align}
S&\xrightarrow{\tau} -\frac{1}{\tau}, & T&\xrightarrow{\tau} 1+\tau.
\label{eqn:A3}
\end{align} 

The finite modular groups $(N\leq5)$ and non-abelian discrete groups are isomorphic to each other \cite{deAdelhartToorop:2011re}. As a result $\Gamma_2\approx S_3$, $\Gamma_3\approx A_4$,$\Gamma_4\approx S_4$,$\Gamma_5\approx A'_5.$

The modular forms $f(\tau)$ of modular level $N$, weight $k$ transform under the action of $\Gamma(N)$ in the following way: 
\begin{equation}f(\gamma \tau)= (c\tau + d )^k f(\tau)
\label{eqn:A4}
\end{equation}

These moduar forms form a linear space of finite dimension i.e. $f_i(\tau)$. For a certain finite modular group, $f_i(\tau)$ transform under a certain unitary representation of that group in the following way:

\begin{equation}f_i(\gamma\tau)= (c\tau + d)^k \rho_{ij}(\gamma) f_j(\tau)
\label{eqn:A5}\end{equation}
where $\rho_{ij}(\gamma)$ is the irreducible representation of that particular group in concern. This particular relation is the foundation of model building of lepton masses and mixing in modular symmetry.

The number of modular forms of weight $2k\hspace{2mm} (k\geq 0)$ and level $N$ is given in the following table (\ref{tab:1}):

\begin{table}[ht]
	\centering
	\begin{tabular}{||c|c|c||}
		\hline
		$N$ &  No. of modular forms& $\Gamma(N)$\\
		\hline
		2 & $k+1$ & $S_3$\\
		\hline
		3 & $2k+1$ & $A_4$\\
		\hline
		4 & $4k+1$ & $S_4$\\
		\hline
		5 & $10k+1$ & $A_5$\\
		\hline
		6 & $12k$ &\\
		\hline
		7& $28k-2$ &\\
		\hline
		
	\end{tabular}
	\caption{No. of modular forms of weight $2k$ \cite{Feruglio:2017spp}.}
	\label{tab:1}
\end{table}
The $\Gamma(3)$ group is isomorphic to the non-abelian discrete symmetric group $A_4$. This group is discussed in the next section.

\subsection{$\Gamma(3)$ modular group}

This is a group of level 3. For the lowest weight i.e. 2, there are three modular forms, $Y(\tau)=(Y_1,Y_2,Y_3)^T.$ These forms are taken as a triplet in $A_4$. They are expressed in the following way \cite{Ding:2019zxk}:
\begin{equation}
\begin{aligned}
& Y_{1}(\tau)=\frac{i}{2\pi}[\frac{\eta'(\frac{\tau}{3})}{\eta(\frac{\tau}{3})}+\frac{\eta'(\frac{\tau+1}{3})}{\eta(\frac{\tau+1}{3})}+\frac{\eta'(\frac{\tau+2}{3})}{\eta(\frac{\tau+2}{3})}-27\frac{\eta'(3\tau)}{\eta(3\tau)}]\\
&    Y_{2}(\tau)=\frac{-i}{\pi}[\frac{\eta'(\frac{\tau}{3})}{\eta(\frac{\tau}{3})}+\omega^2\frac{\eta'(\frac{\tau+1}{3})}{\eta(\frac{\tau+1}{3})}+\omega\frac{\eta'(\frac{\tau+2}{3})}{\eta(\frac{\tau+2}{3})}]\\
& Y_{3}(\tau)=\frac{-i}{\pi}[\frac{\eta'(\frac{\tau}{3})}{\eta(\frac{\tau}{3})}+\omega\frac{\eta'(\frac{\tau+1}{3})}{\eta(\frac{\tau+1}{3})}+\omega^2\frac{\eta'(\frac{\tau+2}{3})}{\eta(\frac{\tau+2}{3})}]
\end{aligned}
\label{eqn:A6}
\end{equation} 

where $\eta(\tau)$ is the Dedekind eta-function and is defined in the following way:
\begin{equation}
\eta(\tau)=q^{\frac{1}{24}}\prod_{n=1}^{\infty}(1-q^{n}), \hspace{1cm} q=e^{2\pi i\tau}.
\label{eqn:A7}
\end{equation}

The eta functions satisfy the equations \begin{equation}
\eta(\tau +1)= \exp^{i\pi/12}\eta(\tau), \hspace{8mm} \eta(-1/\tau)=\sqrt{-i\tau} \eta(\tau)
\label{eqn:A8}
\end{equation}
There is an alternative way of expressing these modular forms. This form of expression is called the $q$-expansion, where $q=\exp(2i\pi\tau)$ and $Y_i$'s are expressed in terms of $q$. Accordingly the modular forms of weight 2 can be written as \cite{Novichkov:2019sqv},
\begin{equation}
\begin{aligned}
	& Y_1(\tau)=1+12q+36q^2+12q^3+.........\\
	& Y_2(\tau)=-6q^{1/3}(1+7q+8q^2+.......)\\
	& Y_3(\tau)=-18q^{2/3}(1+2q+5q^2+......)\\
	\end{aligned}
	\label{eqn:A9}
\end{equation}
The modular forms of higher weights, such as 4, 6 etc. can be obtained from the weight 2 modular forms, by applying the product rules of $A_4$. For example, weight 4 modular forms can be constructed as: $$Y_1^4=((Y_1^2)^3+2Y_2^2Y_3^2), \hspace{3mm} Y_{1'}^4=((Y_3^2)^2+2Y_1^2Y_2^2)$$
$$Y_3^4=\begin{pmatrix}
(Y_1^2)^2-Y_2^2Y_3^2\\(Y_3^2)^2-Y_1^2Y_2^2\\(Y_2^2)^2-Y_1^2Y_3^2
\end{pmatrix}$$
Similarly with the help of these lower weight modular forms we can construct modular forms of higher weights.

\section{$A_4$ symmetry group}
 The group $A_4$ is an even permutation of four objects. It is a symmetry group of the tetrahedron and has (4!/2)=12 elements \cite{Brahmachari:2008fn}. It has four irreducible representations: three are singlets (1,$1',1''$) and a triplet, 3 which is further divided into symmetric ($3_S$) and anti-symmetric ($3_A$) parts. There are two generators of the group, $S$ and $T$, which satisfy the relations $$S^2=(ST)^3=T^3=1$$
 
 The tensor product rules for the irreducible representations are as follows: $$1'\otimes1'=1'', \hspace{4mm} 1'\otimes1''=1, \hspace{4mm} 1''\otimes1''=1'$$ $$3\otimes3=1\oplus1'\oplus1''\oplus3_S\oplus3_A$$
 
 For two triplets, $a=(a_1,a_2,a_3)$ and $b=(b_1,b_2,b_3)$, we can write the following multiplication rules:
 \begin{equation}
 \begin{aligned}
  & 1\hspace{2mm}\equiv(ab)=a_1b_1+a_2b_3+a_3b_2\\
  & 1'\hspace{1mm}\equiv(ab)'=a_3b_3+a_1b_2+a_2b_1\\
  & 1''\equiv(ab)''=a_2b_2+a_1b_3+a_3b_1\\
  & 3_S\equiv(ab)_S=(2a_1b_1-a_2b_3-a_3b_2,\hspace{2mm} 2a_3b_3-a_1b_2-a_2b_1,\hspace{2mm}2a_2b_2-a_1b_3-a_3b_1)\\
  & 3_A\equiv(ab)_A=(a_2b_3-a_3b_2,\hspace{2mm} a_1b_2-a_2b_1, \hspace{2mm} a_1b_3-a_3b_1)
  \end{aligned}
 \end{equation}


\bibliography{doc}

\end{document}